\documentclass[10pt,a4paper]{article}
\usepackage[utf8]{inputenc}
\usepackage{amsmath}
\usepackage{amsfonts}
\usepackage{amssymb}
\usepackage{appendix}
\usepackage{booktabs}
\usepackage[left=2cm,right=2cm,top=2cm,bottom=2cm]{geometry}
\usepackage{graphicx}

\usepackage{dblfloatfix}
\usepackage{enumitem}
\usepackage{multicol}
\usepackage{hyperref}
\usepackage{csquotes}

\usepackage{xcolor}
\hypersetup{
    colorlinks,
    linkcolor={black!50!black},
    citecolor={blue!50!black},
    urlcolor={blue!80!black}
}

\usepackage[round,sort]{natbib}
\newtheorem{defi}{Definition}[section]
\newtheorem{assum}[defi]{Assumption}

\usepackage{sectsty}
\sectionfont{\sffamily\bfseries\large\uppercase}
\subsectionfont{\sffamily\bfseries\normalsize}
\subsubsectionfont{\sffamily\bfseries\small}

\newcommand{\PSSA}{P_{\textrm{SSA}}}
\newcommand{\PSIA}{P_{\textrm{SIA}}}

\begin{document}

\title{Anthropic decision theory for self-locating beliefs}
\author{Stuart Armstrong%
\thanks{Email: \texttt{stuart.armstrong@philosophy.ox.ac.uk}; Corresponding author}}

\date{September 2017}

\maketitle

\begin{abstract}\noindent This paper sets out to resolve how agents ought to act in the Sleeping Beauty problem and various related anthropic (self-locating belief) problems, not through the calculation of anthropic probabilities, but through finding the correct decision to make. It creates an anthropic decision theory (ADT) that decides these problems from a small set of principles. By doing so, it demonstrates that the attitude of agents with regards to each other (selfish or altruistic) changes the decisions they reach, and that it is very important to take this into account. To illustrate ADT, it is then applied to two major anthropic problems and paradoxes, the Presumptuous Philosopher and Doomsday problems, thus resolving some issues about the probability of human extinction.
\end{abstract}


\begin{multicols}{2} 

\section{Introduction}

We cannot have been born on a planet unable to support life. This self-evident fact is an example of anthropic or self-locating reasoning: we cannot be `outside observers' when looking at facts that are connected with our own existence. By realising that we exist, we change our credence of certain things being true or false. But how exactly? Anyone alive is certain to have parents, but what is the probability of siblings?

Different approaches have been used to formalise the impact of anthropics on probability. The two most popular revolve around the `Self-Sampling Assumption' and the `Self-Indication Assumption' \citep{anthbias}. Both of these give a way of computing anthropic probabilities, answering questions such as `Given that I exist and am human, what probability should I assign to there being billions (or trillions) of other humans?'

The two assumptions are incompatible with each other, and give different answers to standard anthropic problems. Nor are they enough to translate probabilities into decisions. Many anthropic problems revolve around identical copies, deciding in the same way as each other, but causal \citep{CDT} and evidential\footnote{
Evidential decision theory (EDT) has been argued to be undefined; see arguments in \url{http://lesswrong.com/lw/e7e/whats_wrong_with_evidential_decision_theory/} and Egan's contention that `fixing' causal decision theory must also `fix' EDT \citep{counter_CDT}, along with Joyce's `fixing' of causal decision theory \citep{regret_CDT}. Anthropic decision theory resembled EDT more than it does CDT, but has distinct foundations.
} \citep{EDT} decision theory differ on whether agents can make use of this fact. And agents using SIA and SSA can end up always making the same decision, while still calculating different probabilities \citep{prob_not_enough}. We are at risk of getting stuck in an intense debate whose solution is still not enough to tell us how to decide.

Hence this paper will sidestep the issue, and will not advocate for one or the other of the anthropic probabilities, indeed arguing that anthropic situations are a distinct setup, where many of the arguments in favour of probability assignment (such as long run frequencies) must fail.

Instead it seeks to directly find the correct decision in anthropic problems. The approach has some precedence: Briggs argues \citep{value_beauty} that SIA-type \emph{decision making} is correct. On the face of it, this still seems an extraordinary ambition -- how can the right decision be made, if the probabilities aren't fully known? It turns out that with a few broad principles, it is possible to decide these problems without using any contentious assumptions about what the agent's credences should be. The aim of this approach is to extend classical decision making (expected utility maximisation) to anthropic situations, without needing to use anthropic probabilities.

This will be illustrated by careful analysis of one of the most famous anthropic problems, the Sleeping Beauty problem \citep{sleeping_beauty}. One of the difficulties with the problem is that it is often underspecified from the decision perspective. The correct decision differs depending on how much Sleeping Beauty considers the welfare of her other copies, and whether they share common goals. Once the decision problem is fully specified, a few principles are enough to decide the different variants of the Sleeping Beauty problem and many other anthropic problems.

That principled approach fails for problems with non-identical agents, so this paper then presents an Anthropic Decision Theory (ADT), which generalises naturally and minimally from the identical agent case. It is not a normative claim that ADT is the ideal decision theory, but a practical claim that following ADT allows certain gains (and extends the solutions to the Sleeping Beauty problem). ADT is based on `self-confirming linkings': beliefs that agents may have about the relationship between their decision and that of the other agents. These linkings are self-confirming, because if believed, they are true. This allows the construction of ADT based on open promises to implement self-confirming linkings in certain situations. Note that ADT is nothing but the Anthropic version of the far more general ``Updateless Decision Theory\footnote{
See for instance \url{https://wiki.lesswrong.com/wiki/Updateless_decision_theory}} and Functional Decision Theory \citep{damascus}.

The last part of the paper applies ADT to two famous anthropic problems: the Presumptuous Philosopher and Doomsday problems, and removes their counter-intuitive components. This is especially relevant, as it removes the extra risk inherent in the Doomsday argument, showing that there is no reason to fear extinction more than what objective probabilities imply \citep{xriskprio}.

\section{Sleeping Beauty Problem}

The Sleeping Beauty \citep{sleeping_beauty} problem is one of the most fundamental in anthropic probability. Many other problems are related to it, such as the absent-minded driver \citep{absent}, the Sailor's Child Problem \citep{FNC}, the incubator and the presumptuous philosopher \citep{anthbias}. In this paper's perspective, all these problems are variants of each other -- the difference being the agent's level of mutual altruism.

\begin{figure*}
\begin{center}
\includegraphics[width=0.9\textwidth]{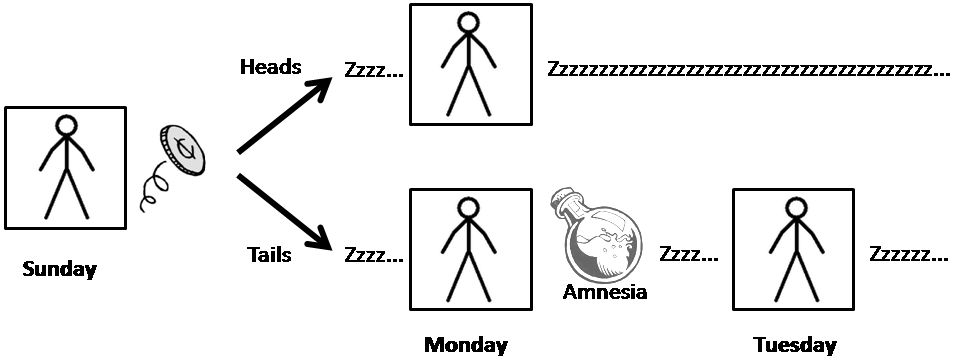}
\end{center}
\caption{The Sleeping Beauty Problem.}
\label{SB:problem}
\end{figure*}

In the standard setup, Sleeping Beauty is put to sleep on Sunday, and awoken again Monday morning, without being told what day it is. She is put to sleep again at the end of the day. A fair coin was tossed before the experiment began. If that coin showed heads, she is never reawakened. If the coin showed tails, she is fed a one-day amnesia potion (so that she does not remember being awake on Monday) and is reawakened on Tuesday, again without being told what day it is. At the end of Tuesday, she is left to sleep for ever (see Figure \ref{SB:problem}).

\begin{figure*}
\begin{center}
\includegraphics[width=0.9\textwidth]{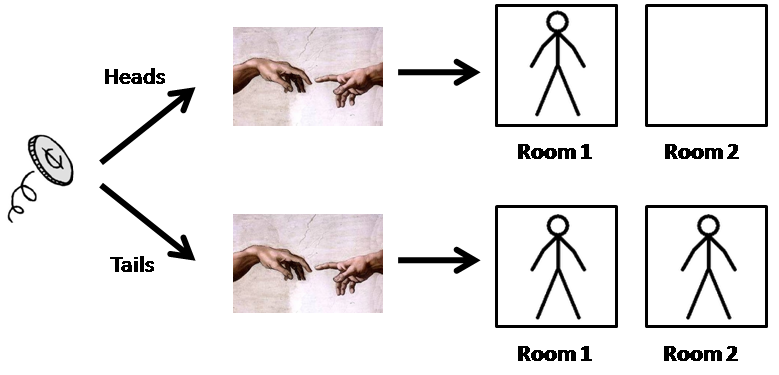}
\end{center}
\caption{Incubator variant of the Sleeping Beauty problem.}
\label{ISB:problem}
\end{figure*}

The incubator variant of the problem \citep{anthbias} has no initial Sleeping Beauty, just one or two copies of her created (in separate, identical rooms), depending on the result of the coin flip (see Figure \ref{ISB:problem}).

The question, traditionally, is what probability a recently awoken/created Sleeping Beauty should give to the outcome of the coin toss or the day of the week/room she's in.

\subsection{Halfers: the Self-Sampling Assumption}

The self-sampling assumption (SSA) relies on the insight that Sleeping Beauty, before being put to sleep on Sunday, expects that she will be awakened in future. Thus her awakening grants her no extra information, and she should continue to give the same credence to the coin flip being heads as she did before, namely $1/2$.

In the case where the coin is tails, there will be two copies of Sleeping Beauty, one on Monday and one on Tuesday, and she will not be able to tell, upon awakening, which copy she is. By the principle of indifference, she should therefore assume that both are equally likely. This leads to:

\begin{assum}[Self-Sampling Assumption]
All other things being equal, an observer should use the standard non-anthropic\footnote{
Non-anthropic means the probability of each world, as judged from an ``outsider'' who doesn't exist within any world (or doesn't use the fact of their existence to update the probabilities).
} probabilities for any world they could be in. For computing their position in any (possible) world with multiple observers, they should assume they are randomly selected from the set of all \emph{actually existent} observers in that world, within their reference class.
\end{assum}

There are some issues with the concept of `reference class' \citep{anthbias}, but here it is enough to set the reference class to be the set of all other Sleeping Beauties woken up in the experiment. We will return to the problems with reference classes in Section \ref{ref:class}, where a fruitful identification is made between the reference class and the people Sleeping Beauty cares about.

The probability calculations are straightforward; upon awakening, Sleeping Beauty will have the following probabilities:
\begin{align*}
\PSSA (\textrm{Heads})& = 1/2 \\
\PSSA (\textrm{Tails})& = 1/2\\
\PSSA (\textrm{Monday}|\textrm{Heads})& = 1\\
\PSSA (\textrm{Tuesday}|\textrm{Heads})& = 0\\
\PSSA (\textrm{Monday}|\textrm{Tails})& = 1/2\\
\PSSA (\textrm{Tuesday}|\textrm{Tails})& = 1/2.
\end{align*}

She then deduces:
\begin{align*}
\PSSA (\textrm{Monday})& = 3/4\\
\PSSA (\textrm{Tuesday})& = 1/4.
\end{align*}

SSA can be called the halfer position, $1/2$ being the probability it gives for Heads. 

\subsection{Thirders: the Self-Indication Assumption}

There is another common way of doing anthropic probability, using the self-indication assumption (SIA). This derives from the insight that being woken up on Monday after heads, being woken up on Monday after tails, and being woken up on Tuesday are all subjectively indistinguishable events. In terms of parallel universes (or many worlds quantum physics \citep{decohere}), in which half of the universes see heads and half see tails, then one third of all of Sleeping Beauty's ``awakening experiences'' would be in the heads universes, and two thirds in the tails universes. This is formalised as:

\begin{assum}[Self-Indication Assumption]
All other things being equal, an observer should reason as if they are randomly selected from the set of all epistemically possible observers.
\end{assum}

The definition of SIA is different from that of Bostrom \citep{anthbias}; what we call SIA he called combined SIA+SSA. We shall stick with the definition above, however, as it is the most common one used in discussions and conferences. Note that there is no mention of reference classes, as SIA is the same for all reference classes\footnote{
As long as the reference class is large enough to contain all the possible states the observer could be in; in this case, the reference class must contain both Monday and Tuesday awakenings and both heads and tails for Monday.
}.

The three following observer situations are equiprobable (each has a non-anthropic probability of $1/2$), and hence SIA gives them equal probabilities $1/3$:

\begin{align*}
\PSIA (\textrm{Monday}\& \textrm{Heads})& = 1/3\\
\PSIA (\textrm{Monday}\& \textrm{Tails})& = 1/3\\
\PSIA (\textrm{Tuesday}\& \textrm{Tails})& = 1/3.
\end{align*}

She then deduces:
\begin{align*}
\PSIA (\textrm{Monday})& = 2/3\\
\PSIA (\textrm{Tuesday})& = 1/3\\
\PSIA (\textrm{Heads})& = 1/3\\
\PSIA (\textrm{Tails})& = 2/3.
\end{align*}

SIA can be called the thirder position, $1/3$ being the probability it gives for Heads.

\section{Are probabilities necessary?}

Much of the discussion around the thirder and halfer positions revolves around reconciling contrary intuitions. Intuitively, it seems correct that the probability of heads or tails, upon awakening, should be $1/2$ (the halfer position). It also seems correct that, upon learning that it is Monday, the probability of heads or tails should be $1/2$ (the thirder position). And, finally, it's intuitively correct that the update rule upon learning that it is Monday makes these two intuitions incompatible.

Then the arguments revolve around rejecting one of these three intuitions (including, in an ingenious argument, the standard Bayesian update rule \citep{SB_again}). This paper argues against all such approaches, by claiming that probability is the wrong tool to be using in this area. This argument can be grounded by analysing how the usual probability arguments break down.

\subsection{Long run frequencies}\label{long:run}

The easiest interpretation of probability is the frequentist one: an event's probability is the limit of the relative frequency of a large number of trials. But it was realised that in the Sleeping Beauty problem, you would get different answers depending on whether you counted the number of times Sleeping Beauty was correct, or the number of trials in which she was correct\footnote{
This is sometimes called the difference between total and average inaccuracy.
} \citep{SB_again,value_beauty}. The first case would imply the thirder position; the second, the halfer.

\subsection{Bayesianism: whose beliefs?}

Bayesian probability grounds probability in states of knowledge or states of belief \citep{bayes_background,prob_theo}. However, this approach has the same problems as frequentism. A well-calibrated Bayesian estimator that assigns a probability of $1/2$ to something, expects to be incorrect half the time \citep{calibration}. But does this means half the times they estimate, or in half the trials? Similarly, the Bayesian estimator will reach different answers depending on whether they first assess the relative probability of universes before assessing their own identity (the halfer position), or vice versa (the thirder position). There are tricky problems distinguishing between ``I expect to see'', and ``I expect someone exactly identical to me to see'', which can make degrees of belief ambiguous.

\subsection{Intuition, identity, and evolution}

Our understanding of probability is connected with our subjective sentiment of identity, the ``subjective credence'' that ``we'' will or won't be surprised by a certain outcome. There is a vast and ancient literature on the concept of identity. It is also clear that personal identity serves a useful evolutionary function, allowing us to successfully plan for the long term rather than leap from a window, confident that someone else will hit the ground\footnote{
See some further thoughts at \url{http://lesswrong.com/lw/k9u/the_useful_definition_of_i/} and \url{http://lesswrong.com/lw/grl/an_attempt_to_dissolve_subjective_expectation_and/}
}.

But evolution does not aid in forging intuitions in situations that did not exist in our evolutionary history. Duplication, merging and amnesia are not standard evolutionary situations, so we should not expect to have good intuitions on them. Maybe if they had been, we would now have a better subjective grasp of the meaning of questions such as ``Who do I expect to be, given that there are two of me, both thinking the same thing and asking the same question?''

\subsection{Dutch Books and Cox's theorem}

There are strong arguments that any formalisation of uncertainty must be done in terms of probability, with Cox's theorem providing the theoretical justification and various Dutch Book arguments giving strong practical reasons to do so \citep{cox,prob_theo}. Both styles of arguments are however compatible with the assignment of different priors. Thus they do actually impose the use of probabilities per se, but require that any process akin to updating probabilities must follow standard Bayesian rules. As shall be seen, though Anthropic Decision Theory doesn't assign probabilities per se, it does respect such update rules\footnote{
To realise this, it suffices to notice that Anthropic Decision Theory can be phrased with various different probability assignments, and then updates them in the traditional way. This demonstrates both the compatibility with Bayesian updates, and the fact that no probability assignment is fundamental.
}.

Some authors \citep{beauty_bets,value_beauty} aim to extend Dutch Book arguments to anthropic situations, arguing for SIA on those grounds. But that argument makes implicit assumptions as to how much each agent values other copies, meaning it fits within some of the frameworks below, but not all of them.

In fact, the failure of Dutch Books to distinguish between SIA and SSA is one of the strongest arguments that conventional probability is the wrong tool to use in anthropic situations. It is not simply a question of the betting odds coming apart from the subjective probabilities; it is a situation where the strongest arguments that compel the need for subjective probabilities no longer apply.

\section{Deciding Sleeping Beauty}

\subsection{How to decide?}

If Sleeping Beauty needs to make a decision after being awakened, then her choice of SSA or SIA is not enough to fix her action. She'd be pretty sure that any copies of her would reach the same decision as her. But can she take this into account? Causal decision theory would tell her to ignore these other decisions (since she is not causing them) and evidential decision theory would say the opposite.

An evidential decision theorist using SSA generally reaches the same decisions as a causal decision theorist using SIA, so neither her anthropic probability theory nor her decision theory can be independently deduced from her actions. How much ``responsibility'' she bears for a joint decision also affects this \citep{prob_not_enough}.

Anthropic decision theory thus cares only about decisions. It makes use only of elements agreed upon by all observers: the non-anthropic probabilities in the experiment, Sleeping Beauty's utility function, the fact that the copies are identical, and the use of classical decision theory in non-anthropic situations.

\subsection{Identical copies, different preferences?}

Copies of Sleeping Beauty are assumed to be identical. Common sense would imply that identical agents have identical preferences. This need not be the case: two nationalists could both believe ``my country right or wrong'', but have different preferences if they lived in different places. If the pronoun ``my'' was replaced by its referent, the two nationalists would no longer be identical. But if their preferences are phrased in pronoun form, identical agents can have distinct preferences.

It is important to know whether copies of Sleeping Beauty value each other's wellbeings. The memory erasure (or the duplication) means that this is no longer self-evident. It is not \emph{prima facie} obvious that the Monday copy would be willing to forgo a chocolate so that her Tuesday copy could have three. It is even less obvious in the incubator variant. Thus we will distinguish four idealised cases.

The first is non-indexical Sleeping Beauty: an agent dedicated to maximising a specific utility function, a single obsessive \emph{cause}. Her life, her pleasure, and that of others, are completely secondary to what she wants to achieve -- be it increasing the number of paperclips, enhancing shareholder value, or awaiting the Messiah.

The second is altruistic total utilitarian Sleeping Beauty. She desires to maximise the wellbeing of all agents, including herself. Her utility function is the sum of the wellbeing of all existent agents.

The third is altruistic average utilitarian Sleeping Beauty, who similarly counts the wellbeing of all equally with her own, but averages over the number of agents in the world\footnote{
Both altruistic Sleeping Beauties have copy-altruistic variants, where Sleeping Beauty only cares about the wellbeing of other Sleeping Beauties.
}.

The above agents have the same preferences as their copies, in both pronoun and referent form. Not so for the selfish Sleeping Beauty, who doesn't care about any other version of herself. This includes spatially distinct copies, or versions of herself after memory erasure. Though these Sleeping Beauties have identically \emph{phrased} preferences -- ``everything to me!'' -- they have distinct preferences: they disagree over which states of the world are better, with each copy wanting to pull the gain in her direction. But selfishness is not actually clearly defined, with two versions of selfishness -- the ``psychological approach'' and the ``physically selfish'' -- resulting in different decisions.

\subsection{Principles of decision}

We define three principles of decision, to allow the Sleeping Beauties to reach decisions without anthropic probabilities. They are presented in decreasing order of intuitiveness; not all three principles are needed in every situation.

\subsubsection{Anthropic diachronic consistency}

The first principle is diachronic consistency: consistency over time. If an agent's preferences change in a predictable manner, then outsiders may be able to exploit this, by selling them something now that they will buy back for more later, or vice versa, and many axiomatic systems seek to avoid losing resources for no purpose in this way. See the independence axiom in the von Neumann-Morgenstern axioms of expected utility \citep{utility}. Non-independent agents show diachronic inconsistency: after partially resolving one of the lotteries, their preferences change. This principle has been useful in other circumstances (even used to deduce the Born rule in quantum mechanics \citep{born}.

We will use consistency in a simple form:

\begin{assum}
Anthropic diachronic consistency (AC):
An agent can commit themselves to a future decision (unless her future copy doesn't remember the commitment, or has different preferences\footnote{
Recall that identical \emph{selfish} copies do not have the same preferences (even if they are the same agent at different moments), so they are not covered by this principle.
}).
\end{assum}

There are many ways to motivate this principle. If an agent would want their future self to take a particular decision, they could take measures to enforce this: leaving money in escrow to be released only under certain conditions, making a public commitment to a certain course of action, or, more classically, tying themselves to the mast so that they can't swim towards the sirens. If they have access to their own motivational structure or their decision algorithms, they could rewire themselves directly. Or more simply, just decide that they would prefer to be a consistent person, and make the effort to become so.

Note that the principle isn't magic: it doesn't avoid the effects of amnesia potions. Sleeping Beauty, on Sunday, could commit her future Monday and Tuesday selves to certain courses of action, but the Monday copy couldn't commit her Tuesday self, as she wouldn't remember that.

AC is the missing piece needed for the Dutch Book arguments of \citep{beauty_bets,value_beauty}, so that the different agents taking the different bets really act as if they were the same person. Even this is not enough, as we shall see: we need to know how exactly the different copies value each other.

\subsubsection{Irrelevant outside details}

In classical decision problems, how the problem came about is considered irrelevant to the decision. This principle extends that irrelevance to anthropic problems.

\begin{assum}
Irrelevant outside details (IOD):
Assume two anthropic problems are identical, in that they have identical agents facing identical choices leading to identical consequences (in terms of their utilities) with identical probabilities. Then the agents will make the same decisions in both problems.
\end{assum}

This principle can be motivated by imagining that Sleeping Beauty is fed an additional amnesia potion, that causes her to forget her past before Monday (but conserves her preferences, knowledge, decision theory, etc...) Then one could approach her when she wakes up offering -- for a price -- to restore her pre-experiment memories\footnote{
In an alternate variant, you could offer to confirm whether or not there had been a past version of herself or not -- or maybe you'd tell her whether Sunday Sleeping Beauty had thought about AC or whether she had wanted to constrain her future copies.
}, just long enough to help her make her current decision. Intuitively, if would seem that she should not pay you for this privilege\footnote{
The past version of Sleeping Beauty certainly wouldn't want her future version to pay to remember her!
}. What extra information could she gain that would affect her decision?

\subsubsection{Irrelevant inside details}

Apart from outside details, classical decision theory also considers many inside details as irrelevant. Only a few details are generally taken to be important: the decisions the agents face, possible outcomes, probabilities, and the utility that each agent has for each outcome.
For copies, there is an extra subtlety: from the outside, it's evident that some outcomes are impossible. The two copies of Sleeping Beauties must make the same decision; there cannot be an outcome where one accepts an offer and the other turns down an identical one\footnote{
Both could choose to accept the gamble with a certain probability -- but they would both name the same probability.
}. We'll call the decisions of the two agents `linked' (see Section \ref{ADT:sec}), and our last principle decrees that when two setups have the same possible linked outcomes, then the decisions of the agents must be the same.

\begin{assum}
Irrelevant inside details (IID):
If two situations have the same setup, the same possible linked decisions and the same utility outcomes for each agent in every possible linked decision, and all agents are aware of these facts, then agents should make the same decisions in both situations.
\end{assum}

A causal decision theorist could object to this principle: the linked decision concept seems to implicitly lock in an evidential decision theory. But one could argue that this is the behaviour that any agents (including causal decision agents) would want to prescribe to others, including to future copies. Since this principle is only needed for Sleeping Beauties that are selfish, one could conceive of it as a deal: it would be advantageous for all selfish agents to follow this principle, and they would desire it to be universally implemented, if possible. Some of issues will discussed more at length in Section \ref{ADT:sec}.

To see how all these principles apply, we will now look at a specific decision problem.

\subsection{The decision problem}

In order to transform the Sleeping Beauty problem into a decision problem, assume that \emph{every time} she is awoken, she is offered a coupon that pays out $\pounds1$ if the coin fell tails. She must then decide at what price she is willing to buy that coupon; for the rest of the paper, we will assume her utility is linear in money (non-indexicals being allowed to invest their income directly in their cause, with linear returns).

The decisions made by different types of agents are labelled ``SIA''-like or ``SSA''-like, since some will behave as we would intuitively expect a thirder or halfer agent to behave.

\subsubsection{Non-indexical Sleeping Beauty (SIA-like)}

A non-indexical Sleeping Beauty, before she is put to sleep the first time, will reason:

\begin{displayquote}
``In the tails world, future copies of me will be offered the same deal twice. Since they cannot distinguish between themselves, they will accept (or reject) the same deal twice. Any profit they make will be dedicated to our common cause, so from my perspective, profits (and losses) will be doubled in the tails world. If my future copies' decision algorithm told them to buy the coupon for $\pounds x$, there would be an expected $\pounds 1/2(2(-x + 1) + 1(-x + 0)) = \pounds (1-3/2x)$ going towards our goal. Hence I would want them to buy for $x<\pounds 2/3$, or refrained from buying for $x>\pounds 2/3$.''
\end{displayquote}

Then by AC, she would change her decision procedure to ensure her future copies will follow this prescribed behaviour, buying below $\pounds 2/3$ and refraining above.

\subsubsection{Non-indexical Sleeping Beauty without the initial Beauty (SIA-like)}

In some variants of the Sleeping Beauty problem (such as the sailor's child \citep{sailor_child} and the incubator), there is no initial Sleeping Beauty to make decisions for her future copies. Thus AC is not enough to resolve the decision problem. Then we need to invoke IOD. By this principle, the incubator variant is unchanged if we started with an initial Sleeping Beauty that was then copied (or not) to create the beings in the experiment. That initial Beauty would follow the same reasoning as the Sunday Beauty in the standard variant. Hence by AC, these copies would buy for $x<\pounds 2/3$, and refrain from buying for $x>\pounds 2/3$. IOD reduces the non-indexical incubator Sleeping Beauty to the standard non-indexical Sleeping Beauty.

\subsubsection{Altruistic total utilitarian Sleeping Beauty (SIA-like)}

Non-indexical agents (who don't care about anyone's wellbeing) can be distinguished from altruistic agents. The non-indexical agent only cares about the monetary aspect of a bet in as much as the money can be used to further their preferred outcome; the altruist cares about the monetary aspect of a bet in as much as the money provides personal welfare to each winner.

However, an altruistic total utilitarian will have the same preferences over the possible outcomes in a Sleeping Beauty situation\footnote{
It has been pointed out that an entirely altruistic agent would also care about the profit of whomever it was that offered her the bet! Hence we'll postulate that the bet is offered by a non-conscious machine, and that its value is ultimately created or destroyed (rather than being given to, or taken from, someone).
}: the impact of outcomes in the tails world is doubled, as any gain/loss happens twice, to two different agents (or two different agent-slices), and the altruist adds up the effect of each gain/loss. It is not hard to see that the altruistic total utilitarian Sleeping Beauty will make the same decisions as the non-indexical one.

\subsubsection{Copy-altruistic total utilitarian Sleeping Beauty (SIA-like)}

The above argument does not require that Sleeping Beauty be entirely altruistic, only that she be altruistic towards all her copies. Thus she may have selfish personal preferences (``I prefer to have this chocolate bar, rather than letting Cinderella get it''), as long as these are not towards her copies (``I'm indifferent as to whether I or Sleeping Beauty II gets the chocolate bar''). And then she will make the same decision in this betting problem as if she was entirely altruistic.

\subsubsection{Copy-altruistic average utilitarian Sleeping Beauty (SSA-like)}

In Section \ref{long:run}, we mentioned how different ways of combining long run results (total correct versus average correct) led to SSA or SIA being considered correct. Similarly, average and total utilitarians will behave differently in the Sleeping Beauty problem.

Consider the reasoning of an initial, copy-altruistic average utilitarian Sleeping Beauty that considers her Monday and Tuesday copies as two different agents (since they won't remember each other):

\begin{displayquote}
``If the various Sleeping Beauties decide to pay $\pounds x$ for the coupon, they will make $-\pounds x$ in the heads world. In the tails world, they will each make $\pounds (1-x)$ each, so an average of $\pounds (1-x)$. This creates an expected utility of $\pounds 1/2(-x+(1-x))=\pounds (1/2-x)$, so I would want them to decide to buy the coupon for any price less than $\pounds 1/2$, and refrain from buying it at any price more than $\pounds 1/2$.''
\end{displayquote}

And hence by AC, this will be the behaviour the agents will follow. Thus they would be behaving \emph{as if} they were following SSA odds, and putting equal probability on the heads versus tails world. As above, for the incubator variant, we can use IOD to get the same behaviour (``average utilitarian between copies'' is more intuitive in the incubator variant).

\subsubsection{Altruistic average utilitarians and reference classes}\label{ref:class}

Standard SSA has a problem with reference classes. For instance, the larger the reference class becomes, the more the results of SSA in small situations become similar to SIA, a puzzling effect as the bounds of the reference class can seem somewhat arbitrary.

The same situation arises with an altruistic average utilitarian Sleeping Beauty (rather than a copy-altruistic one). Assume there are $\Omega$ other individuals in the population that Sleeping Beauty cares about. Then if the copies accept the coupon, the averaged gain in the tails world is $2(1-x)/(2+\Omega)$, since the total population is $2+\Omega$. The average loss in the heads world is $-x/(1+\Omega)$, since the total population is $1+\Omega$.

If $\Omega$ happens to be zero (because the agents are copy-altruistic, or because there is nobody else in the universe), then the situation proceeds as above, with the indifference price of the coupon being $\pounds 1/2$. However, as $\Omega$ increases, the indifference price of the coupon rises. If $\Omega$ is very large, then $2(1-x)/(2+\Omega) - x/(1+\Omega)$ is approximately $(2-3x)/\Omega$. Thus the agent's indifference price will be just below $\pounds (2/3)$ -- the same as in the total utilitarian case, a behaviour that looks ``SIA-ish''.

Hence one of the paradoxes of SSA (that changing the reference class changes the agent's probabilities), is, in this context, closely analogous to one of the difficulties with average utilitarianism: that changing the population one cares about changes one's decisions.

\subsubsection{Selfish Sleeping Beauty (SIA or SSA-like)}

In all above examples, the goals of one Sleeping Beauty were always in accordance with the goals of her copies. But what happens when this fails? When the different versions are entirely selfish towards each other? Easy to understand in the incubator variant (the different created copies may feel no mutual loyalty), it can also happen in the standard Sleeping Beauty problem if she simply doesn't value a time slice of herself that she doesn't remember (or that doesn't remember her).

And here we hit the problem of defining selfishness. It is actually considerably more subtle to define selfishness than altruism\footnote{
In the author's view, the confusion illustrated by this section shows that selfishness is not a stable concept: those agents capable of changing their definitions of selfishness will be motivated to do so.
}. This is because even a selfish agent is assumed to value a sufficiently close future version of herself -- if not, the agent cannot be said to reach decisions, or even to be an agent. But different ways of drawing the line result in different preferences. One definition is the brute-physical view: that physical continuity defines identity \citep{personal_identity}. Another definition emphasises psychological continuity \citep{constitute_identity}. One can deduce preferences from these definitions\footnote{
In the author's view, the deduction should run the other way: from preferences, one can infer a useful concept of identity, that can be different for different agents with different preferences.
}: physical continuity would imply that agent doesn't value the other copy of themselves in the tails world, nor do they value a heads world in which they didn't exist. The psychological approach would similarly fail to value the other copy in the tails world, but would value the -- psychologically identical -- version of themselves in the heads world.

The distinction can be made clearer referring to the copies as ``Sleeping Beauty'' (SB) and ``Snow White'' (SW), otherwise identical agents with distinct labels. The physically selfish SB values only SB, and makes a distinction between ``I exist, with no copies'' and ``An identical copy of me exists, but I don't\footnote{
This view is counterfactually relevant in the tails world; there is also an issue of a ``Rawlesian veil of ignorance'' \citep{veil} in the heads wold, but it is unclear that such a veil can be extended to not knowing one's own existence.
}''. In contrast, the psychologically selfish SB values only SB in the tails world, but values whichever agent exists in the heads world\footnote{
We now take these to be the definition of physical and psychological selfishness, deliberately avoiding the issues of personal identity that gave rise to them.
}.

Thus we consider three worlds, with different expected rewards in each, given that the agents pay $\pounds X$ for the coupon (the rewards are expressed for SB; those for SW are symmetric). See the outcomes in Table \ref{SW:SN}.

\begin{table*}
\begin{center}
\begin{tabular}{|c|c|c|c|c|}
\hline
World & Total util. & Average util. & Phy. Selfish & Psy. Selfish\\
\hline
Heads, SB exists & $-X$ & $-X$ & $-X$ & $-X$\\
Heads, SW exists & $-X$ & $-X$ & $0$ & $-X$\\
Tails (both exist) & $2-2X$ & $1-X$ & $1-X$ & $1-X$\\
\hline
\end{tabular}
\end{center}
\caption{Sleeping Beauty and Snow White's utility in depending on their preferences and the outcome}
\label{SW:SN}
\end{table*}

The non-anthropic probabilities of these three worlds are $1/4$, $1/4$, and $1/2$, respectively. Notice first that psychological selfishness results in exactly the same utilities as the average utilitarian. Invoking IID means that they will give the same SSA-style decision.

The expected utility of the physically selfish agent, however, is half of that of the total utilitarian agent (and the non-indexical agent). Since utility can be rescaled, IID can then be invoked to show that they will make the same SIA-style decision. Note that the factor of two is relevant in cases where agents have preferences that are a mix of these ``pure'' preferences; the total utilitarian pulls towards the thirder position more than the physically selfish agent does.

\subsection{Summary of results}

We have five categories of agents\footnote{
We could extend to more exotic categories, such as Sleeping Anti-Beauty (\url{http://lesswrong.com/lw/4e0/sleeping_antibeauty_and_the_presumptuous/}) where the two copies in the tails world have antagonistic motivations. Psychological selfish Sleeping Beauty can correctly be seen as a $50-50$ uncertainty between the altruistic total utilitarians Sleeping Beauty and the Sleeping Anti-Beauty.
}, and they follow two different types of decisions (SIA-like and SSA-like). In the Sleeping Beauty problem (and in more general problems), the categories decompose:

\begin{enumerate}
\item Non-indexical agents who will follow SIA-like odds.
\item (Copy-)Altruistic total utilitarians who will follow SIA-like odds.
\item (Copy-)Altruistic average utilitarians who will follow SSA-like odds.
\item ``Psychologically'' selfish agents who will follow SSA-like odds.
\item ``Physically'' selfish agents who will follow SIA-like odds.
\end{enumerate}

For the standard Sleeping Beauty problem, the first three results are derived from anthropic diachronic consistency. The same result can be established for the incubator variants using irrelevant outside details principle. The selfish results, however, need to make use of the irrelevant inside details principle.

\section{Beauties vary in the eye of the beholders: anthropic decision theory}\label{ADT:sec}

The previous section has strong claims to being normative -- the three principles can be defended (to slightly different degrees) as being intuitive and necessary. They did however require that each Beauty be strictly identical to each other. Given this fact, the Beauties would reach the same decision in the same circumstances -- and they would know this fact, and know that they knew this fact, and so on, making this fact \emph{common knowledge} \citep{reason_know}.

In this situation, there are strong arguments for superrationality \citep{superrat} -- for using this common knowledge to reach common goal. But this breaks down if the agents are not identical (or are not known to be identical).

Superrationality is often used to argue that agents in the famous prisoner's dilemma (PD) should mutually cooperate. And, similarly, identical agents in PD can cooperate using similar reasoning about common knowledge -- and anthropic agents can be seen as similarly ``cooperating'' to buy the multiple coupons in the tails world. But cooperation between non-symmetric agents is tricky: they can disagree as to who should get the majority of the gains from trade, making the process into a negotiation rather than a simple decision\footnote{
The issue doesn't arise for identical agents, because one cannot gain (in expectation) at the expense of the other if they pursue identical strategies.
}.

It is for this reason that this section is not normative: Anthropic Decision Theory needs to be associated with a ``fair'' (or at least agreed-upon) method for dividing gains, before it is complete. Since such division is not relevant to the Sleeping Beauty problem, we will simply assume that there is a single way of doing this, and elide the problem.

\subsection{Cooperation among diversity}

Some agents can cooperate with each other even if they are distinct and non-interacting. Agents using causal decision theory \citep{CDT} will not cooperate with anyone in these types of situations, as the different agent's decisions are not causally connected. Similarly, it would be foolish for any agent to cooperate with a causal decision theorist, as they would not return the ``favour''. In contrast, two evidential decision makers could cooperate with each other if they had the belief that it was likely that they would be reaching the same decision.

More generally, consider some mystical brand X that compels the bearer to cooperate with any other bearer of X. This allows two X-bearers to cooperate, and hence reach the same decisions as two identical Sleeping Beauties in the problem above.

This does not guarantee general cooperation: X-bearers need not cooperate with bearers of an otherwise identical brand Y. But the question is: if you were in an anthropic situation, and had the opportunity to take the brand X (and Y, and Z, and so on for all other mystic group-cooperation-forcing brands), should you?

From a causal decision perspective, the answer is clearly no. But if there were some common ancestor to both you and the other agent(s), then they would clearly want all agents to take every cooperation brand they could get their hands on.

There are no mystical cooperation brands in reality; but, assuming that people are capable of \emph{precommitting} themselves to certain courses of action\footnote{
Or of programming automated decision making machines to do so.
} \citep{hitchhiker}, there are ways of reaching the same state of cooperation.

\subsection{Self-confirming linkings}

When agents reach decisions, define a linking between them as:

\begin{defi}[Linking]
A linking between the decisions of two agents is a restriction on the possible joint decisions that the two agents could reach.
\end{defi}

For instance, the knowledge that both agents are identical and will decide similarly, gives such a linking. The claim that both agents will reach opposite decisions is another such linking. But note what happens for the first case, in both the Prisoner's Dilemma and the Sleeping Beauty problem: if both agents believe they will reach the same decision, they will act on this and reach individual decisions... that will indeed be the same\footnote{
Mutual cooperation for the Prisoner's Dilemma.
}. Whereas if the agents believe they will reach opposite decisions, they will act on this and reach individual decisions... that will be identical to each other\footnote{
Mutual defection for the Prisoner's Dilemma.
}. Thus, if both agents believe the first linking, it will be true; if they believe the second, it will be false.

So we define a general self-confirming linking\footnote{
The self-confirming linking has some formal similarities to ``self-committing'' and ``self-revealing'' utterances in game theory \citep{multi_agent}.
}:

\begin{defi}[Self-confirming linking]
A self-confirming linking between two agents is a linking such that if both agents believe it, they will reach decisions that make the linking true.
\end{defi}

Thus, choosing to accept a self-confirming linking as a fact about the world, plays a similar role to accepting the mystical brand X: it allows cooperation between agents that do so. And since it is self-confirming, it is actually true. Note that causal decision theory can be seen as accepting a self-confirming linking along the lines of ``the other agent will maximise their utility as if we reached decisions independently''.

\subsection{Anthropic Decision Theory}

With this definition, defining Anthropic Decision Theory is simple. An ADT agent is simply one that is willing to search for self-confirming linkings\footnote{
There could be multiple self-confirming linkings possible -- for instance, causal decision theory has one. It's clear that in many situations, there are self-confirming linkings that are strictly better, for all agents, than the causal decision theory one. However, there might not be a Pareto-dominant option that all agents could agree on: this brings us back to the issue of negotiations, which we're currently eliding.
} and implement them -- accept them as a premise of reasoning:

\begin{defi}[Anthropic decision theory (ADT)]
An ADT agent is an agent that would implement a self-confirming linking with any agent that would do the same. It would then maximises its expected utility, conditional on that linking, and using the standard non-anthropic probabilities of the various worlds.
\end{defi}

It's not hard to see that if the Sleeping Beauties in the previous chapter were all ADT agents, they would all reach the ``correct'' decisions computed there. Hence ADT is an extension to non-identical agents of the decision-centric approach of the previous section.

As an illustration of this extension, consider the non-indexical sleeping Beauty problem, but where each agent has an independent $50\%$ chance of valuing a coupon at $5$ pence more than its face value -- maybe the colour is attractive (since the agents are non-indexical, rather than total utilitarian, they derive no utility from having the other agent appreciate the coupon).

Then the following linking is always self-confirming: If the two agents are offered a coupon priced below $\pounds 2/3$, or above $\pounds (2/3+5/100)$, then they will make the same decision.

Thus non-identical agents can `cooperate' to `solve' anthropic problems\footnote{
One would like to extend this result to coupons priced between $\pounds 2/3$ and $\pounds (2/3+5/100)$.
}, `defecting' only on the range $\pounds 2/3$ to $\pounds (2/3+5/100)$ (where neither agent will buy the coupon). Note that on that range, the agents will have the same behaviour, but this will only be true if they don't assume this fact when making their decision: the linking ``Both agents will output the same decision'' is not self-confirming on that range.

The obvious generalisation of this is to situations where the linking is likely but not certain. But there are subtleties there, connected with the negotiation problem; this is beyond the scope of this paper\footnote{
To give a flavour of the difficulties: the linking previously described is self-confirming, while the linking ``both agents output the same decision'' is self-confirming with a certain probability. Depending on whether or not they have the ``$+\pounds 5/100$'' valuation, different agents will prefer one or the other of the two linkings, or certain intermediate ones. And are the agents supposed to decide before or after realising their extra valuation?
}.

\subsubsection{Infinite ethics}

SIA and SSA break down when there are infinite numbers of observers -- indeed SIA fails when the \emph{expected} number of observers is infinite\footnote{
\url{http://lesswrong.com/lw/fg7/sia_fears_expected_infinity/}
}.

ADT, on the other hand, does not intrinsically have problems with infinite numbers of observers. If the agents' utilities are bounded, or if there are a finite number of options, then ADT functions exactly as in the finite agent cases. Thus ADT can apply to some situations where SIA and SSA both break down.

\section{Applying Anthropic Decision Theory}

This section applies anthropic decision theory to two major problems in anthropic probability: the Presumptuous Philosopher and the Doomsday argument. A later paper will look into other anthropic problems, such as the UN++ and the Adam and Eve problem \citep{adameve}.

\subsection{Presumptuous Philosopher}

The Presumptuous Philosopher was introduced by Nick Bostrom \citep{anthbias} to point out the absurdities in SIA. In the setup, the universe either has a trillion observers, or a trillion trillion trillion observers, and physics is indifferent as to which one is correct. Some physicists are about to do an experiment to determine the correct universe, until a presumptuous philosopher runs up to them, claiming that his SIA probability makes the larger one nearly certainly the correct one. In fact, he will accept bets at a trillion trillion to one odds that he is in the larger universe, repeatedly defying even strong experimental evidence with his huge SIA probability correction.

Note that the argument doesn't work if there is only one presumptuous philosopher in either universe. For then, the number of other observers doesn't matter: the philosopher will find each universe equiprobable (he's a trillion trillion times more likely to exist in the larger universe, but a trillion trillion times less likely to be the presumptuous philosopher; these effects exactly cancel). So the expected number of presumptuous philosophers in the larger universe needs to be a trillion trillion times larger than in the smaller universe. Let's take the simplest case: there is one presumptuous philosopher in the smaller universe, and a trillion trillion of them in the larger.

What does ADT have to say about this problem? Implicitly, the philosopher is often understood to be selfish towards all, including towards any copies of himself (similarly, Sleeping Beauty is often implicitly assumed to be non-indexical or copy-total altruistic, which may explain the diverging intuitions on the two problems) -- or, at least, the issue of altruism is not seen as relevant.

If the philosopher is selfish, ADT reduces to SSA-type behaviour: the philosopher will correctly deduce that in the larger universe, the other trillion trillion philosophers or so will have their decision linked with his. However, he doesn't care about them: any benefits that accrue to them are not of his concern, and so if he correctly guesses that he resides in the larger universe, he will accrue a single benefit. Hence there will be no paradox: he will bet at 1:1 odds of residing in either universe.

If the philosopher is an altruistic total utilitarian, on the other hand, he will accept bets at odds of a trillion trillion to one of residing in the larger universe. But this no longer counter-intuitive (or at least, no more counter-intuitive than maximising expected utility for large utilities and very small probabilities): the other presumptuous philosophers will make the same bet, so in the larger universe, total profits and losses will be multiplied by a trillion trillion. And since the philosopher is altruistic, the impact on his own utility is multiplied by a trillion trillion in the large universe, making his bets rational.

It might be fair to ask what would happen if some of the philosophers were altruistic while others were selfish. How would the two interact; would the altruistic philosopher incorrectly believe his own decision was somehow `benefiting' the selfish ones? Not at all. The decisions of the altruistic and selfish philosophers are not linked: they both use ADT, but because they have very different utilities, their linking is not self-confirming: they would reach different decisions even if they all believed in the linking.

\subsection{Doomsday argument}

The Doomsday argument \citep{doom} is an important and famous argument in anthropic reasoning. Based on SSA's preference for `smaller' universes, it implies that there is a high probability of the human race becoming extinct within a few generations -- much higher than objective factors would imply. In a very simplified form, the argument states that we are much more likely to be one of a small group than one of a large group, so it is much more likely that there are fewer humans across time than many. Since we're pretty solid on the number of humans in the past, this implies a smaller future: hence a doomsday for our species at some point soon.

Under SIA, the argument goes away \citep{doomSIA}, so it seems that ADT must behave oddly: selfish and altruistic agents would give different probabilities about the extinction of the human race! But recall that it is decisions that ADT cares about, not probabilities -- so can we phrase a reasonable Doomsday \emph{decision}?

SSA-like behaviour comes from psychologically selfish or average utilitarian agents. Focusing on the average utilitarians, assume that there is an event $X$ with probability $p$ which will ensure humanity's survival. If that happens, then there will be a total of $\Omega$ humans; otherwise, the human population will cap at $\omega < \Omega$.

Assume there is another event $Y$ which is independent of $X$, but also has a probability of $p$. The agent must choose between a coupon that pays out $\pounds 1$ on $X$, and one that pays out $\pounds 1$ on $Y$.

The expected utility of the $X$ coupon is simply $p/\Omega$, since $\Omega$ will be the population if $X$ happens. The expected utility of the Y coupon is a bit more complicated, since with probability $p$, its expected value is $p/\Omega$, and with probability $(1-p)$, it is $p/\omega$. Putting all these together gives an expected value of $p[p/\Omega + (1-p)/\omega]$. That quantity is greater than $p/\Omega$, thus the agent will choose the coupon paying on $Y$.

This reproduces the original Doomsday Argument: two events of equal probability, $X$ and $Y$, and yet, because $X$ is connected with human survival while $Y$ is independent of it, the agent chooses the coupon paying on $Y$. And the larger the population gets -- $\Omega$ relative to $\omega$ -- the more pronounced this effect becomes.
This \emph{looks} like a fear of doom (and would be, if a total utilitarian made the same decision), but is clearly just a quirk of the average utilitarian's preference system.

\section{Conclusion}

Anthropic decision theory is a new way of dealing with anthropic (self-locating belief) problems, focused exclusively on finding the correct decision to make, rather than the correct probabilities to assign. It deals successfully with many classical anthropic puzzles, and resolves various paradoxes such as the Presumptuous Philosopher and the Doomsday argument, in ways that don't clash with intuition.

In many situations, ADT is moreover a consequence of simple principles of Anthropic diachronic consistency, Irrelevant outside details and Irrelevant inside details. In order to apply it fully, some facts became very important, such as whether the agents were altruistic or selfish, total or average utilitarians -- facts that are generally not relevant in most non-anthropic problems. But when making a decision in anthropic situations, they matter tremendously: the difference between SIA-like and SSA-like behaviour is down to how copies value each other's goals.

\section*{Acknowledgments}
This paper has been a long time in preparation, and I want to gratefully acknowledge the help of many people, too numerous to mention or even remember, from the Future of Humanity Institute, the Machine Intelligence Research Institute, philosophers working in anthropic probability, and the online community of Less Wrong.

\bibliography{../ref}

\end{multicols}
\end{document}